\documentclass[aps,prb,twocolumn,floatfix,showpacs,showkeys]{revtex4-1}
\usepackage{amsfonts}
\usepackage{amsmath}
\usepackage{amssymb}
\usepackage{graphicx}
\usepackage{color}

\begin{document}

\title{Metallic nanograins: spatially nonuniform pairing induced by
quantum confinement}


\author{M. D. Croitoru$^{1,3}$}
\author{A. A. Shanenko$^2$}
\author{C. C. Kaun$^3$}
\author{F. M. Peeters$^2$}
\affiliation{$^1$Institut f\"{u}r Theoretische Physik III,
Universit\"{a}t Bayreuth, 95440 Bayreuth, Germany}
\affiliation{$^2$Departement Fysica, Universiteit Antwerpen,
Groenenborgerlaan 171, B-2020 Antwerpen, Belgium}
\affiliation{$^3$Research Center for Applied Sciences, Academia
Sinica, 11529 Taipei, Taiwan}

\keywords{metallic nanograins, nonuniform pairing, superconducting
correlations, matrix elements} \pacs{74.20.Fg, 74.78.Na}

\begin{abstract}
It is well-known that the formation of discrete electron levels
strongly influences the pairing in metallic nanograins. Here we
focus on another effect of quantum confinement in superconducting
grains that was not studied previously, i.e., spatially nonuniform
pairing. This effect is very significant when single-electron levels
form bunches and/or a kind of shell structure: in highly symmetric
grains the order parameter can exhibit variations with position by
an order of magnitude. Nonuniform pairing is closely related to a
quantum-confinement induced modification of the pairing-interaction
matrix elements and size-dependent pinning of the chemical potential
to groups of degenerate or nearly degenerate levels. For
illustration we consider spherical metallic nanograins. We show that
the relevant matrix elements are as a rule enhanced in the presence
of quantum confinement, which favors spatial variations of the order
parameter, compensating the corresponding energy cost. The
size-dependent pinning of the chemical potential further increases
the spatial variation of the pair condensate. The role of nonuniform
pairing is smaller in less symmetric confining geometries and/or in
the presence of disorder. However, it always remains of importance
when the energy spacing between discrete electron levels $\delta$ is
approaching the scale of the bulk gap $\Delta_B$, i.e., $\delta
> 0.1$-$0.2\,\Delta_B$.
\end{abstract}

\date{\today }
\maketitle

\section{Introduction}
\label{sec1}

Quantum confinement plays a fundamental role in superconductors with
nanoscale dimensions. Interplay of quantum confinement and pairing
correlations results in important qualitative changes in the
superconductor characteristics.~\cite{parm,blat,mush,smit, brau,
perali,sier,glad,yuzb,kres01,kres02,sh01,sh02,sh03,sh04,sh05,gar}
Because of technological reasons quasi-0D superconducting structures
(i.e., ensembles of small grains) were the first where this
interplay was investigated experimentally. Initial attempts by
Giaever and Zeller at the end of 60s used tunneling studies on large
ensembles of superconducting particles.~\cite{gaiv01} Since that
time most of the studies on superconducting correlations in grains
were performed with grain powders~\cite{li01,li02} or on films made
of crystalline granules separated by amorphous inter-granular
space.~\cite{bose01, bose02} In the pioneering work of Ralph et
al.~\cite{ralph01, ralph02} the discrete electron spectrum was
measured for a single grain. Their technique (single-electron
tunneling spectroscopy) enabled them for the first time to probe
superconducting correlations in an individual ${\rm Al}$ grain. Very
recently, STM was used to detect the superconducting gap of an
isolated ultra-small lead grain deposited onto a silicon substrate
(see e.g. Refs.~\onlinecite{bose03,bose04}). These advances opened
new prospects to examine superconductivity in individual metallic
nanograins with unprecedented detail, e.g., to investigate the
influence of the confinement on the superconducting correlations.

The main feature of a superconducting nanograin that makes them
different from a bulk superconductor is the formation of discrete
electron levels with average energy spacing $\delta \approx 2\pi^2
\hbar^2/(m k_F V)$, with $k_F$ the bulk Fermi wave number and $V$
the system volume. It can be of the same order as the bulk gap
$\Delta_B$, or even larger in the case of ultra-small nanograins.
Therefore, size-quantization of the electron spectrum can have a
substantial impact on the basic superconducting characteristics of
such quasi-0D superconducting systems.

The understanding of the fundamental properties of superconducting
correlations in low-dimensional structures, in particular in
isolated metallic grains, has experienced a remarkable development
in the last two decades. Theoretical aspects, which have attracted
the most attention are the following. The problem of the breakdown
of BCS superconductivity in ultra-small metallic grains was
addressed in several papers.~\cite{delft,smit,mat}  The effect of
the shell structure in the single-electron spectrum on the
superconducting correlations was pointed out for
nanograins~\cite{glad,gar} and ultrasmall metallic
clusters.~\cite{kres01,kres02} The ground state properties of the
BCS pairing Hamiltonian of ultra-small grains were considered beyond
the mean-field approximation using the Richardson exact
solution.~\cite{sier,glad,yuzb}

A spatially uniform pairing was assumed in these and other works
and, as a consequence, the matrix elements of the pairing
interaction were taken independent of the relevant single-electron
quantum numbers, i.e., they were set to $-g/V$, with $g>0$ the
coupling constant and $V$ the volume.~\cite{note1} This is, say, a
bulk-like approximation recovered when the single-electron wave
functions are taken as plane waves. However, the translational
invariance is broken in nanograins, which leads to a
position-dependent order parameter. As a result, the pairing gap
becomes strongly dependent on the relevant quantum numbers, which is
directly related to a confinement-induced modification (as compared
to $-g/V$) of the matrix elements controlling the scattering of the
time reversed states. Another important issue is that
single-electron levels can form bunches and even a kind of shell
structure in symmetric confining geometries. In this case the
chemical potential $\mu$ can be pinned to a group of nearly
degenerate or degenerate levels. This is of importance because the
density of states in the vicinity of $\mu$ strongly influences the
superconducting correlations. In other words, such a pinning plays
the role of a filter that selects the contribution of a particular
single-electron shell (or of a group of close levels) to the
superconducting order parameter. Such a contribution is, as a rule,
spatially nonuniform.

The aim of the present paper is to investigate effects related to a
spatially nonuniform pairing in metallic nanograins, which was not
studied in previous publications. For illustrative purposes we
consider metallic spherical nanograins, where the spatial dependence
of the superconducting condensate is pronounced (the order parameter
can vary with position by an order of magnitude). In less symmetric
confining geometries and/or in the presence of disorder spatial
variations of the order parameter are reduced. However, our study
implies that nonuniform pairing remains of importance when the
interlevel spacing $\delta$ is approaching the scale of the order of
the bulk gap $\Delta_B$. Any remaining grouping of single-electron
levels, that is always present in real samples, even strengthens the
effect of interest. We work in the mean-field approximation and,
thus, stay in the regime $\delta \lesssim \Delta_B$. Below we
consider ${\rm Sn}$ and ${\rm Al}$ with $\Delta_B = 0.616$ and
$0.25\,{\rm meV}$, respectively~(for the parameters used below).
Using the above values of $\Delta_B$, we find that the mean-field
approach is valid for $D > 6$-$8\,{\rm nm}$, with $D$ the sphere
diameter.

Our paper is organized as follows. In Sec.~\ref{sec2}, we outline
the formalism how to obtain a self-consistent solution to the
problem. In Sec.~\ref{sec3}, we present our numerical results. In
particular, in Sec.~\ref{sec3a} we investigate the effects of
quantum confinement on pairing correlations through the
modifications of the matrix elements of the pairing interaction and
the size-dependent pinning of $\mu$ to single-electron shells.
Sec.~\ref{sec3b} is focused on a spatial distribution of the pair
condensate and its relation to modifications of the matrix elements
and the size-dependent pinning of $\mu$. In Sec.~\ref{sec3c} we
discuss the interplay of Andreev reflection with quantum
confinement, resulting in the formation of Andreev-type states and
significant dependence of the pairing gaps on the relevant quantum
numbers. A short summary and discussion are given in
Sec.~\ref{sec4}.

\section{Formalism}
\label{sec2}

The reduction of the system to the nanometer scale leads to the
formation of a discrete electron spectrum. Moreover, in the presence
of quantum confinement, the translational invariance of the system
is broken, and the superconducting order parameter is position
dependent, i.e., $\Delta=\Delta({\bf r})$. For the mean-field
treatment of such a situation, it is appropriate to use the
Bogoliubov-de Gennes (BdG) equations,~\cite{bogl,degen} which can be
written as
\begin{subequations}
\begin{align}
&E_i|u_i\rangle =\widehat{H}_e|u_{i}\rangle +
                                             {\widehat{\Delta}}
|v_i\rangle,  \label{bdgA} \\
&E_i |v_i\rangle ={\widehat{\Delta}}^{\ast}|u_{i}\rangle
-\widehat{H}_e^{\ast}|v_i\rangle, \label{bdgB}
\end{align}
\end{subequations}
where $E_i$ stands for the Bogoliubov-quasiparticle (bogolon)
energy, $\widehat{\Delta}=\Delta(\widehat{\bf r})$~(with
$\widehat{\bf r}$ the position operator) and the single-electron
Hamiltonian is referred to the chemical potential $\mu$, i.e.,
\begin{equation}
\widehat{H}_e({\bf r})=\frac{\widehat{\bf p}^2}{2m_e}+V
(\widehat{\bf r})-\mu.  \label{He}
\end{equation}
We remark that any magnetic effects are beyond the scope of the
present paper. For simplicity, the confining interaction $V({\bf
r})$ is taken as zero inside the specimen and infinite outside:
$V({\bf r})=V_B\;\vartheta(R-\rho)$ with the barrier potential
$V_B\to\infty$~($R=D/2$ and $\rho$ is the radial coordinate for
the spherical confining geometry).

As a mean-field approach, the BdG equations should be solved in
a self-consistent manner
\begin{equation}
\Delta({\bf r})=g\sum_i\langle{\bf r}|u_i\rangle\langle v_i|
{\bf r}\rangle \tanh(\frac{\beta E_i}{2}), \label{delta_1}
\end{equation}
where $g>0$ is the coupling constant for the effective
electron-electron interaction approximated by the delta-function
potential, i.e., $\langle {\bf r}, {\bf r}'|\Phi| {\bf r},{\bf r}'
\rangle=-g\delta({\bf r}-{\bf r}')$. The sum in Eq.~(\ref{delta_1})
runs over the states with the single-electron energy
\begin{equation}
\xi_i=\bigl[\langle u_i|\widehat{H}_e|u_i\rangle +\langle v_i|
\widehat{H}_e|v_i\rangle\bigr] \in [-\hbar\omega_D, \hbar\omega_D],
\label{ksi}
\end{equation}
with $\omega_D$ the Debye frequency. As is known, the solution of
the BdG equations has two branches: $(i,+)$ and $(i,-)$~(see Ref.~
\onlinecite{swid}) for which we have $E_{i,+}>0$ and $E_{i,-}<0$.
The sum in Eq.~(\ref{delta_1}) should be taken over the physical
states [the $(i,+)$ branch], i.e., $E_i = E_{i,+}$.

For a given mean electron density $n_e$ the chemical potential $\mu$
is determined from
\begin{equation}
n_e=\frac{2}{V}\sum_i \bigl[f_i\langle u_i|u_i\rangle+(1-f_i)
\langle v_i|v_i\rangle\bigr], \label{dens}
\end{equation}
with $V = \frac{4}{3}\pi R^3$ the volume of the spherical grain. For
conventional superconductors the energy gap is typically much
smaller than the chemical potential. As a result, $\mu$ stays nearly
the same when passing from the normal state to the superconducting
one.~\cite{degen} Therefore, one can solve Eq.~(\ref{dens}) in the
absence of superconducting order ($\Delta({\bf r})=0$).

In a spherical nanograin, because of symmetry reasons, the order
parameter depends only on the radial coordinate, i.e.,
$\Delta=\Delta(\rho)$. Therefore the pseudospinor in the
particle-hole space can be characterized by the quantum numbers of
the angular momentum, i.e., ($l,m$). The angular part of the
pseudospinor $\Psi_i$ is given by the spherical harmonics $Y_{lm}
(\theta,\varphi)$ in polar coordinates $(\rho,\theta,\varphi)$,
i.e.,
\begin{equation}
\langle {\bf r}|\Psi_i\rangle=Y_{lm}(\theta,\varphi)
\left(\begin{array}{c}
u_{jl}(\rho)\\
v_{jl}(\rho)
\end{array}
\right),
\label{factor}
\end{equation}
where $i=\{j,l,m\}$, with $j$ the radial quantum number associated
with the quantum-confinement boundary conditions
\begin{equation}
u_{jl}(\rho)\vert_{\rho=R}=v_{jl}(\rho)\vert_{\rho=R}=0.
\label{boundary}
\end{equation}

To solve the BdG equations ~(\ref{bdgA}) and (\ref{bdgB})
numerically, $u_{jl} (\rho)$ and $v_{jl}(\rho)$ are expanded in the
eigenfunctions of the single-electron Hamiltonian $\widehat{H}_e$
[see Eq.~(\ref{He})]. In addition, iterations should be invoked, to
account for the self-consistency relation given by
Eq.~(\ref{delta_1}). This program is significantly simplified by
keeping only the pairing of the time-reversed states,~\cite{and}
which is a standard approximation for the problem of superconducting
correlations in nanograins. In the framework of the BdG equations
this can be done through the so-called Anderson approximate solution
for which the particle- and hole-like wave functions are assumed to
be proportional to the single-electron wave function. It means that
\begin{equation}
u_{jl}(\rho) = {\cal U}_{jl}\,\chi_{jl}(\rho),\; u_{jl}(\rho) = {\cal
V}_{jl}\,\chi_{jl}(\rho),
\label{basic}
\end{equation}
with the radial part of the single-electron wave function given by
\begin{equation}
\chi_{jl}(\rho)=\frac{\sqrt{2}}{R^{3/2} j_{l+1}(\alpha_{jl})} j_l
(\alpha_{jl}\frac{\rho}{R}),  \label{sef}
\end{equation}
with $j_l(x)$ the $l$-order spherical Bessel function of the first
kind and $\alpha_{jl}$ its $j$-node. The coefficients ${\cal
U}_{jl}$ and ${\cal V}_{jl}$ (taken as real) obey the standard
constraint (see, e.g., Refs.~\onlinecite{ket})
\begin{equation}
{\cal U}^2_{jl} + {\cal V}^2_{jl}=1.
\label{constraint}
\end{equation}
Then, inserting Eq.~(\ref{basic}) into Eqs.~(\ref{bdgA}) and
(\ref{bdgB}) we find the following set of coupled equations (here
$E_{jlm}=E_{jl}$ and $\xi_{jlm}=\xi_{jl}$):
\begin{subequations}
\begin{align}
[E_{jl}-\xi_{jl}]~{\cal U}_{jl}=\Delta_{jl}\,{\cal V}_{jl},
\label{matr1}\\
[E_{jl}+\xi_{jl}]~{\cal V}_{jl}=\Delta_{jl}\,{\cal U}_{jl},
\label{matr2}
\end{align}
\end{subequations}
with
\begin{equation}
\Delta_{jl}=\int\limits_{0}^{R}{\rm d}\rho \,\rho^2\,
\chi^2_{jl}(\rho)\Delta(\rho)\label{Delta_matr}
\end{equation}
and
\begin{equation}
\xi_{jl}=\frac{\hbar^2}{2m_e}\frac{\alpha_{jl}^2}{R^2}-\mu.
\label{xi_jl}
\end{equation}
A nontrivial physical solution of Eqs.~(\ref{matr1}) and
(\ref{matr2}) exists only when
\begin{equation}
E_{jl}= \sqrt{\xi_{jl}^2 +\Delta^2_{jl}}. \label{spectrumA}
\end{equation}

The Anderson prescription about the pairing of the time-reversed
states allows one to rephrase the self-consistency relation [see
Eq.~(\ref{delta_1})] as follows:
\begin{equation}
\Delta_{j'l'}= -\sum\limits_{jl}(2l+1)\;
\frac{M_{j'l',jl}\;\Delta_{jl}}{2\sqrt{\xi^2_{jl}+\Delta^2_{jl}}}
\tanh(\frac{\beta E_{jl}}{2}), \label{selfAnder}
\end{equation}
where
\begin{eqnarray}
M_{j'l',jl}=-\frac{g}{4\pi}\int\limits_0^R\!\!{\rm d}\rho\, \rho^2
~\chi_{j'l'}^2(\rho)~\chi_{jl}^2(\rho).&\nonumber \label{M_mat}
\end{eqnarray}
To derive Eq.~(\ref{selfAnder}), one should keep in mind the
property of the spherical harmonics $\sum_{m=-l}^{l}
\;|Y_{lm}(\theta,\varphi)|^2=\frac{2l+1}{4\pi}$. We remark that
$M_{j'l',jl}$ is nothing else but the pairing-interaction matrix
element $\langle i',\bar{i'}|\Phi|i,\bar{i}\rangle$~(with
$\bar{i}=\{j,l,-m\}$) averaged over the states with $m=-l,\ldots l$
and $m'=-l',\ldots l'$, i.e.,
$$
M_{j'l',jl}=\frac{1}{(2l'+1)(2l+1)}\sum\limits_{m'=-l'}^{l'}
\sum\limits_{m=-l}^{l}\langle i',\bar{i'}|\Phi|i,\bar{i}\rangle.
$$

As seen from Eq.~(\ref{Delta_matr}), a spatially uniform order
parameter means that the pairing gaps $\Delta_{jl}$ do not depend on
the quantum numbers $j$ and $l$. This is compatible with
Eq.~(\ref{selfAnder}) only when $M_{j'l',jl}$ does not depend on
$j'$ and $l'$. According to the definition given by
Eq.~(\ref{M_mat}), we have $M_{j'l',jl}=M_{jl,j'l'}$ and, so, if
$M_{j'l',jl}$ does not depend on $j',l'$, it does not depend on
$j,l$ either. So, we arrive at the standard simplified approach of
investigating the pairing correlations in metallic grains (see the
discussion in the Introduction). Below we show that the spatial
dependence of the order parameter can not be ignored in
superconducting nanograins, which implies significant variations of
the matrix elements and pairing gaps with the relevant quantum
numbers. After a numerical solution of Eq.~(\ref{selfAnder}), the
position-dependent order parameter can be calculated from
\begin{equation}
\Delta(\rho)=\sum\limits_{jl} \Delta^{(jl)}(\rho),
\label{orderpar}
\end{equation}
with the shell-dependent contribution $\Delta^{(jl)}(\rho)$ given by
\begin{equation}
\Delta^{(jl)}(\rho)= \frac{g}{8\pi}\,(2l+1)\,\frac{\chi_{jl}^2
(\rho)\,\Delta_{jl}}{\sqrt{\xi^2_{jl}+\Delta^2_{jl}}}
\tanh(\frac{\beta E_{jl}}{2}). \label{orderpartial}
\end{equation}

\section{Discussion of results}
\label{sec3}
\subsection{Enhanced intrashell matrix elements and quantum-size
pinning of the chemical potential}
\label{sec3a}
\begin{figure}[tbp]
\resizebox{0.68 \columnwidth}{!}{\rotatebox{0}{
\includegraphics{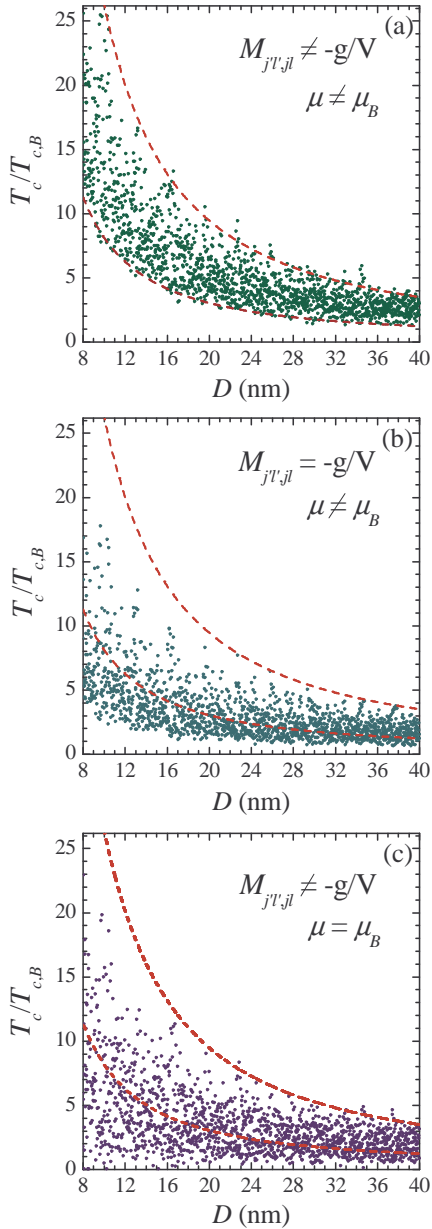}}}
\caption{Critical temperature versus the grain diameter as
calculated for: (a) $M_{j'l',jl} \neq -g/V$ and $\mu \neq \mu_B$;
(b) $M_{j'l',jl} = -g/V$ and $\mu \neq \mu_B$; and (c) $M_{j'l',jl}
\neq -g/V$ and $\mu = \mu_B$. The dashed curves in (a) show
approximate lower and upper boundaries for the quantum-size
oscillations of $T_c$, both curves represent the same dependence
$T_c/T_{c,B}=1+ a(D/D_0)^{3/2}$, with $D_0 = 50\,{\rm nm}$ and
$a=1$~(the lower boundary) and $a=3.5$~(the upper one). The same
curves are also given in (b) and (c), for comparison.} \label{fig01}
\end{figure}

Numerical calculations were performed with the set of parameters
typical for tin~\cite{degen,fett}: $\hbar \omega_D/k_B=195~{\rm K}$,
$gN(0)=0.25$, with $N(0)$ the bulk density of states at the Fermi
level~(we use the bulk electron density $n_e = 148\,{\rm nm}^{-3}$,
see, e.g., Ref.~\onlinecite{ash}).

Figure~\ref{fig01}(a) shows the critical temperature (in units of
the bulk critical temperature $T_{c,B}$) versus the nanograin
diameter $D$ as calculated from Eq.~(\ref{selfAnder}) when the
matrix elements of the electron-electron interaction and the
size-dependent variation of the chemical potential have been fully
taken into account. Results in Fig.~\ref{fig01} are presented for a
step $\Delta R=0.01~{\rm nm}$. For each radius the critical
temperature was defined as the temperature above which the
spatially-averaged order parameter $\langle\Delta(\rho)\rangle$
becomes smaller than $0.01$ of its value at $T=0$. Our numerical
results exhibit two features typical of the size-dependent pairing
characteristics in high-quality superconducting nanograins and
nuclei. First, we observe an overall increase of $T_c$ with
decreasing $D$~(it is very pronounced due to the highly-symmetric
confining geometry). Second, $T_c$ oscillates wildly with $D$. This
oscillatory behavior can be understood in the following way. The
pair correlations are nonzero only for the states within a finite
range (the Debye window) around the chemical potential $\mu$.
Moreover, the main contribution to the sum in Eq.~(\ref{selfAnder})
comes from the states in the very vicinity of the Fermi level,
because in this case the expression $\Delta_{jl}/\sqrt{\xi^2_{jl}+
\Delta^2_{jl}} \simeq 1 $ $(\xi_{jl}\simeq 0)$. When varying the
nanograin size, the number of states in the Debye window changes.
The smaller the diameter, the smaller the number of relevant states
contributing to the pairing characteristics and, as a result, the
more significant is such a change. This change is not monotonous but
rather oscillating due to a permanent competition between incoming
and outcoming states. As a consequence, all basic pairing
characteristics, e.g., $T_c$ and pairing gaps $\Delta_{jl}$, exhibit
quantum-size oscillations. It is not only typical of nanograins with
superconducting correlations (see, e.g., the recent
paper~\cite{bose04}) but it is also present in superconducting
nanowires~\cite{sh01,sh02,sh03,sh04,sh05} and
nanofilms.~\cite{guo,eom} Such oscillations are pronounced for small
diameters/thicknesses but decay with increasing the characteristic
size so that $T_c$ approaches the bulk critical temperature
$T_{c,B}$~(for our parameter $T_{c,B}=4.01\,{\rm K}$). It is
interesting to note that the overall increase of $T_c$ with
decreasing $D$ in Fig.~\ref{fig01}(a) is similar to a size-dependent
enhancement of the pairing gap in nuclei, where it is proportional
to $1/\sqrt{A}$~(see, e.g., Ref.~\onlinecite{sat}), with $A$ the
number of nucleons. In particular, the two dashed curves in
Fig.~\ref{fig01}(a) show approximate upper and lower boundaries for
$T_c$, highlighting the magnitude of the quantum-size oscillations:
both curves represent the same dependence, i.e., $T_c/T_{c,B}= 1 +
a\,(D_0/D)^{3/2}$, with $D_0 = 50\,{\rm nm}$ and $a=1$ and $3.5$ for
the lower and upper boundaries, respectively [$(D_0/D)^{3/2} \propto
N^{-1/2}_e$, with $N_e=n_eV$ the number of electrons]. We remark
that real samples exhibit inevitable shape and size fluctuations
that affect the high-degeneracy of single-electron levels. Hence,
measurements on an ensemble of nanograins will significantly smooth
the quantum-size oscillations in the critical temperature and reduce
its overall enhancement with decreasing nanograin size (see, also,
Sec.~\ref{sec4}). For instance, in experimentally fabricated tin
nanograins of a semi-spherical shape the observed enhancement of the
excitation gap over its bulk value is about~\cite{bose04} $60\%$ for
the particle heights $\approx 10$-$20\,{\rm nm}$. This is
significantly smaller than the enhancement of $T_c$ shown in
Fig.~\ref{fig01}(a). However, detailed investigations of the
enhancement of $T_c$ in superconducting nanograins is beyond the
scope of our present paper. Here we are interested in a spatially
nonuniform distribution of the pair condensate which is of
importance even in the presence of shape and size fluctuations and
disorder (see the discussion in Sec.~\ref{sec4}).

In order to outline the role of the matrix elements $M_{j'l' ,jl}$
[see Eq.~(\ref{M_mat})] of the electron-electron interaction we also
show what happens when the true matrix elements are simply replaced
by those of the bulk-like form: $M_{j'l',jl}=-g/V$, which is what is
usually done when investigating the superconducting correlations in
nanograins. The results are displayed in Fig.~\ref{fig01}(b) and, as
seen, the difference with respect to Fig.~\ref{fig01}(a) is
significant. To simplify the comparison, we show also in
Fig.~\ref{fig01}(b) two solid curves that represent the
radius-dependent upper and lower values of $T_c$ from
Fig.~\ref{fig01}(a).

\begin{table}[ht]
\caption{Matrix elements $M_{j'l',jl}= M_{jl,j'l'}$ in units of
$-g/V$ calculated at $D=7.1\,{\rm nm}$ for quantum numbers such that
$\xi_{j'l'},\xi_{jl} < \hbar \omega_D$:} \centering
\begin{tabular}{|c|c|c|c|c|}
\hline $M_{j'l',jl}$ & $j'$ & $l'$ & $j$ & $l$ \\ \hline
   10.62      &  31  & 11   & 31  &  11 \\
   1.9        &  31  & 11   & 23  &  29 \\
   1.33       &  31  & 11   & 19  &  39 \\
   0.64       &  31  & 11   & 8   & 71  \\
   0.41       &  31  & 11   & 1   & 101 \\
   4.71       &  23  & 29   & 23  &  29 \\
   1.7        &  23  & 29   & 19  &  39 \\
   0.7        &  23  & 29   & 8   & 71  \\
   0.43       &  23  & 29   & 1   & 101 \\
   3.72       &  19  & 39   & 19  &  39 \\
   0.77       &  19  & 39   & 8   &  71 \\
   0.46       &  19  & 39   & 1   & 101 \\
   2.69       &  8   & 71   & 8   &  71 \\
   0.69       &  8   &  71  & 1   & 101 \\
   3.61       &  1   & 101  & 1   & 101 \\
\hline
\end{tabular}
\end{table}

To clarify the physical reason why using the true matrix elements
leads to significant deviations from the results found for
$M_{j'l',jl}=-g/V$, we show in Table I the numerical values of
$M_{j'l',jl}$~(calculated in units of $-g/V$) for $D=14.2\,{\rm
nm}$~(only the states within the Debye window are given). As seen,
the diagonal (intrashell) matrix elements $M_{jl,jl}$ are strongly
enhanced as compared to $-g/V$. However, the matrix elements
controlling the scattering of the time reversed states between
different shells (intershell) are often decreased in absolute value
with respect to $-g/V$. So, the question arises why the
superconducting correlations are enhanced for the true matrix
elements? The point is that the intershell interactions are of less
importance due to a size-dependent pinning of the chemical potential
to the groups of degenerate or nearly degenerate levels (shells can
be often close to each other in energy), see the next paragraph.
When $\mu$ is pinned to a particular shell, then the single-electron
energy measured from $\mu$ is zero for the states from this shell.
These states make a major contribution to superconducting
correlations unless diameters are not large enough $D <
20$-$30\,{\rm nm}$, in other words, the number of contributing
shells is less than $10$-$15$. In this case the superconducting
correlations are nearly determined by the pairing gap $\Delta_{jl}$
associated with the shell pinned to $\mu$. From
Eq.~(\ref{selfAnder}) it is seen that $\Delta_{jl}$ for the states
with $\xi_{jl}=0$ is mainly governed by the intrashell matrix
element $M_{jl,jl}$. For instance, when ignoring the contribution of
all other states one simply obtains (at $T=0$)
$$
\Delta_{jl} \approx - (l+\frac{1}{2})\,M_{jl,jl}.
$$
When the diameter increases beyond $20$-$30\,{\rm nm}$, then the
intershell matrix elements approach $-g/V$ while the intrashell
matrix elements are still significantly different from the bulk-like
behavior. However, the role of the states with $\xi_{jl}=0$ is
becoming less and less important for larger diameters due to the
presence of larger and larger number of shells making a contribution
to the pairing correlations. As a consequence, the difference
between the data in Figs.~\ref{fig01}(a) and (b) decreases when
approaching $D=35$-$40\,{\rm nm}$, together with the amplitude of
the quantum-size oscillations of $T_c$.

\begin{figure}[t]
\resizebox{0.68\columnwidth}{!}{\rotatebox{0}{
\includegraphics{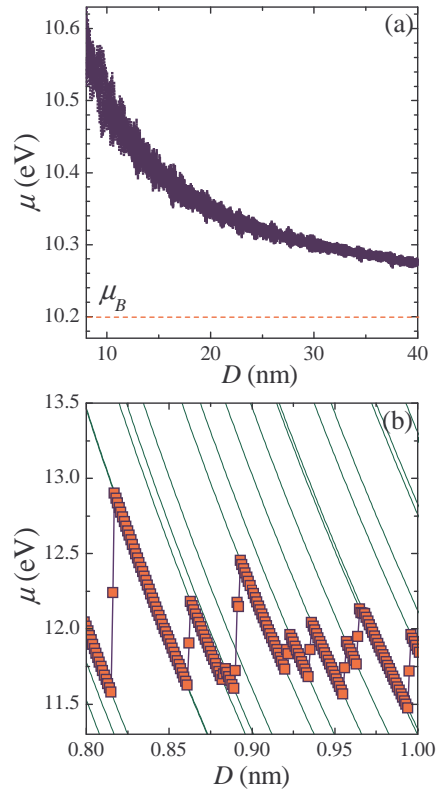}}}
\caption{(a) Size variations of the chemical potential, accompanied
by an overall shift of $\mu$ to upper values with decreasing $D$.
(b) Details of the quantum-size pinning of $\mu$~(filled squares) to
the single-electron levels (solid curves), small diameters are shown
for simplicity.} \label{fig02}
\end{figure}

\begin{figure*}[tbp]
\resizebox{1.8\columnwidth}{!}{\rotatebox{0}{
\includegraphics{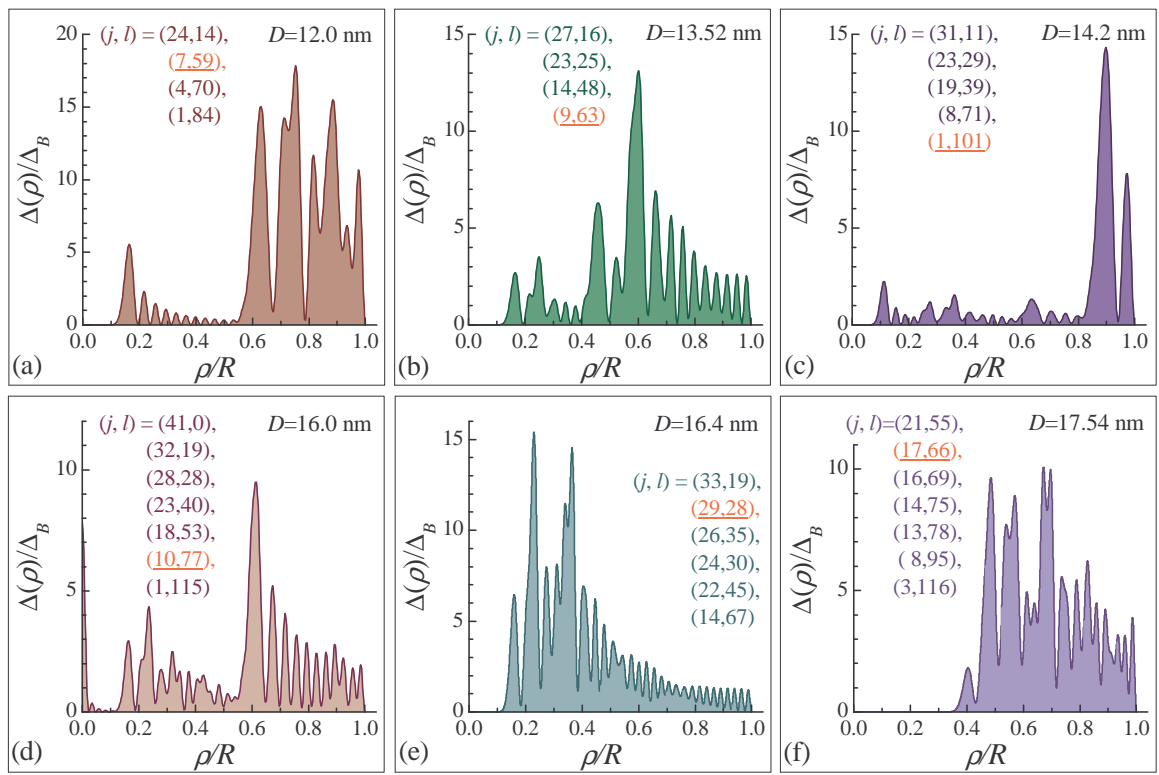}}}
\caption{Spatial distribution of the pair condensate in spherical
nanograins: $\Delta(\rho)$~(calculated at $T=0$) versus $\rho$ for
diameters $D=12\,{\rm nm}$~(a), $13.52\,{\rm nm}$~(b), $14.2\,{\rm
nm}$~(c), $16\,{\rm nm}$~(d), $16.4\,{\rm nm}$~(e) and $17.54\,{\rm
nm}$~(f).} \label{fig03}
\end{figure*}

In the fully self-consistent scheme the chemical potential is
determined in such a way that the mean electron density $n_e$ is
constant [see Eq.~(\ref{dens})]. However, size-dependent variations
of $\mu$ are of importance not only because they simply prevent the
mean electron density from deviations. In fact, such deviations are
almost insignificant: our calculations for $\mu=\mu_B$ show that
$n_e$ decreases by a few percent when $D$ reduces to $10$-$20\,{\rm
nm}$. A more interesting thing is that the size-dependent variations
of $\mu$ have a pronounced effect on the superconducting
correlations. In particular, this can be seen from
Fig.~\ref{fig02}(c), where $T_c$ is calculated for the true matrix
elements and $\mu=\mu_B$. What is the reason for this suppression of
$T_c$? In the presence of the formation of strongly degenerate
electron levels or bunches of electron levels with almost negligible
spacing between them, the chemical potential lies mostly at the
highest partly-filled degenerate level (see, e.g.,
Refs.~\onlinecite{kres01,kres02}). Pairing correlations are
significant only within the Debye window around the chemical
potential $\mu$ and are strongest~\cite{delft} exactly at $\mu$.
Hence, when $\mu$ is pinned to a shell level, this favors the
pairing correlations at this level and, in turn, through the
self-consistency relation, favors the pairing correlations at
neighboring shells. In other words, if the level to which the
chemical potential is pinned is highly degenerate than the phase
space for the strongest pair scattering is enlarged and,
consequently, the system gains in interaction energy and, as a
result, superconducting correlations are strongly enhanced. It is
different when $\mu$ is not pinned to a shell, which is mostly the
case for a constant chemical potential, e.g., for $\mu=\mu_B$. Here
the relevant shells entering the Debye window are as a rule
specified by $\xi_{jl}\not=0$ and, so, their contributions are
diminished.

The above discussion is further illustrated by our numerical results
for $\mu$ in Fig.~\ref{fig02}. As seen from panel (a), when keeping
the electron density of the system constant, $\mu$ slightly shifts
systematically up with decreasing $D$ and exhibits size-dependent
oscillations, as seen from Fig.~\ref{fig02}(a). These oscillations
are a signature of the size-dependent pinning of $\mu$ to groups of
degenerate or nearly degenerate single-electron levels. This is
clearly seen from Fig.~\ref{fig02}(b), where variations of
$\mu$~(filled squares) are plotted versus $D$ together with the
single-electron energies measure from the band bottom, i.e.,
$\frac{\hbar^2}{2m_e} \frac{\alpha^2_{jl}}{R^2}$~(solid curves). For
the sake of simple illustration, panel (b) shows the data for
extremely small diameters, where the energy spacing between the
shell levels is pronounced and, as a result, the size-dependent
oscillations of $\mu$ are not so wild as it happens for higher
diameters. As follows from Fig.~\ref{fig02}(b) $\mu$ is pinned to a
shell level in most cases, which, as mentioned above, represents
incomplete shells. Sometimes $\mu$ can be found between two
neighboring shell levels, which corresponds to the case of a fully
occupied lower shell.

\subsection{Spatially nonuniform pair condensate}
\label{sec3b}

In the previous paragraph we considered the effect of quantum
confinement on pairing correlations through the matrix elements and
quantum-size pinning of $\mu$. As discussed at the end of
Sec.~\ref{sec2}, a framework which incorporates both issues appears
to be only consistent when the position-dependent superconducting
order parameter is taken into consideration. Thus, our results
discussed in the previous section suggest that the spatial
variations of $\Delta(\rho)$ will be pronounced even in nanograins
with diameters up to $D=20$-$30\,{\rm nm}$. However, it is usually
argued that spatial variations of $\Delta(\rho)$ cost significant
extra energy and, so, they are strongly suppressed when $D \ll \xi$,
with $\xi$ the bulk coherence length (see, for instance,
Ref.~\onlinecite{mat}). In addition, $D$ should be larger than
$\lambda_F$: in practice, $k_FD \sim 10$ is assumed to be sufficient
to ignore any spatial dependence of the order
parameter.~\cite{kres01,kres02} For typical metallic parameters
$k_FD \sim 200$-$400$ for $D=10$-$20\,{\rm nm}$ and this is the
reason why the spatial dependence of the order parameter was ignored
in most papers on superconducting correlations in nanograins. To go
in a more detail on this point, we below discuss our numerical
results on $\Delta(\rho)$.

In Fig.~\ref{fig03} the radial dependence of the superconducting
order parameter is shown as calculated from Eq.~(\ref{orderpar}) for
$D=12\,{\rm nm}$~(a), $13.52\,{\rm nm}$~(b), $14.2\,{\rm nm}$~(c),
$16\,{\rm nm}$~(d), $16.4\,{\rm nm}$~(e) and $17.54\,{\rm nm}$~(f).
The shells making a contribution to the superconducting correlations
are also displayed in each panel, and the quantum numbers of the
shell level pinned to $\mu$ are underlined. As seen, we in general
have a nonuniform distribution of the pair condensate for diameters
$D=10$-$20\,{\rm nm}$, which is in agreement with our expectations.
For example, let us consider the results plotted in panel (c). Here
$\mu$ is pinned to the shell level $(l,j)=(101,1)$ and, so,
single-electron states with $j=1$ and $l=101$ make a major
contribution to $\Delta (\rho)$, which results in a significant
enhancement of the order parameter next to the edge, i.e., for
$\rho/R=0.9$-$1.0$. The profile of this enhancement is determined by
the radial wave function $\chi^2_{1,101}(\rho)$ with two pronounced
local maxima ($\Delta/\Delta_B = 14.3$ and $7.2$ at $\rho/R=0.9$ and
$0.97$, respectively) and one node (recall that $j$ is the number of
the nodes of the radial wave function). All the other shells
displayed in Fig.~\ref{fig03}(c) are specified by $\xi_{jl} \not=0$
and, as a result, their contributions is much less significant. The
local maximum $\Delta(\rho)/\Delta_B = 2.3$ at $\rho/R = 0.1$ is due
to states $(j,l)=(31,11)$. The shells with $(j,l)=(23,29)$ and
$(19,29)$ are responsible for local enhancements of the order
parameter up to $1.5$-$2.0\Delta_B$ at $\rho/R=0.27$ and $0.36$,
respectively. At last, the shell $(8,71)$ produces the local maximum
at $\rho/R=0.64$. In general, the larger the angular momentum, the
larger the values of $\rho/R$ at which the corresponding
single-electron states have an effect on the profile of
$\Delta(\rho)$.

It is worth noting that typically, the order parameter is strongly
suppressed in the center ($\rho=0$) except of rare cases when states
with zero angular momentum contribute to the pairing correlations.
One such example is given in Fig.~\ref{fig03}(d), where a narrow
pick can be seen at $\rho=0$ due to the contribution of the shell
with $(j,l)=(41,0)$.

\begin{figure*}[tbp]
\resizebox{1.8\columnwidth}{!}{\rotatebox{0}{
\includegraphics{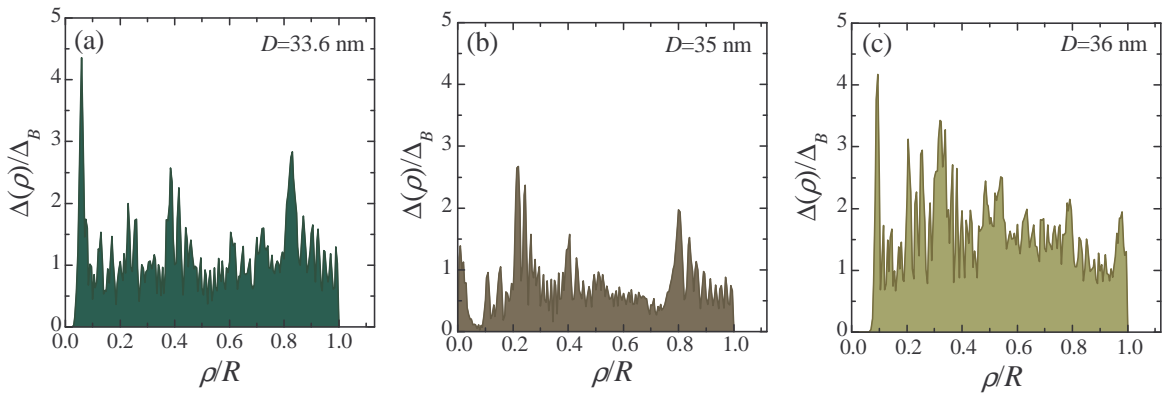}}}
\caption{The order parameter $\Delta(\rho)$ for sufficiently large
diameters $D=33.6\,{\rm nm}$~(a), $D=35\,{\rm nm}$~(b) and $D=36\,
{\rm nm}$~(c).} \label{fig04}
\end{figure*}

From Fig.~\ref{fig03} it follows that the radial distribution of the
pair condensate remains strongly nonuniform even for $D \approx
20\,{\rm nm}$. We would like to note that when selecting concrete
values of $D$ for Fig.~\ref{fig03}, we did not even take diameters
for which $T_c$ is close to the upper dashed curve in
Fig.~\ref{fig01}(a). In the case of a strong enhancement of $T_c$
the radial distribution of the pair condensate is as a rule strongly
nonuniform. The points selected for Fig.~\ref{fig03} are mainly in a
vicinity of the lower dashed curve in panel (a) of Fig.~\ref{fig01}:
for $D=13.52,\,14.2,\,16$ and $16.4\,{\rm nm}$ we have $T_c/T_{c,B}
=6.41,\,6.48,\,4.082$ and $3.78$, respectively. However, even in
this case the order parameter can vary with position by an order of
magnitude. Spatial variations of $\Delta(\rho)$ are significantly
relaxed only when $D$ approaches $30$-$40\,{\rm nm}$, as seen from
Fig.~\ref{fig04}.

For our parameters $k_F = 16.4\,{\rm nm}$ and, so, we obtain $k_FD
\approx 300$ for $D\approx 20\,{\rm nm}$. Hence, the criterion $k_FD
\gg 1$ is not very useful in order to estimate the effect of spatial
variations of the pair condensate. Based on our numerical study, we
would like to suggest another criterion related to a more sensitive
energy scale, which in the superconducting state is governed by the
bulk pairing gap $\Delta_B$. The spatial distribution of the order
parameter is always strongly inhomogeneous when $\delta \sim
\Delta_B$~(here it is even better to replace $\Delta_B$ by the
size-dependent pairing gap). The spatial variations decay with a
decrease in the ratio of the mean interlevel spacing to the bulk
order parameter, i.e., $\delta/\Delta_B$, and our numerical results
suggest that such variations are significantly reduced only when
$\delta/\Delta_B < 0.05$-$0.1$~(recall that effects of a magnetic
field are beyond the scope of our paper). For ${\rm Sn}$ spherical
superconducting grains this regime is achieved when $D >
40$-$50\,{\rm nm}$~(note that $\delta\approx 2\pi^2 \hbar^2/(m k_F
V)$ underestimates the intershell spacing for spherical confining
potential). Despite that our results are for a highly symmetric
confining geometry, we can expect that the order parameter will be
always spatially nonuniform for $\delta/\Delta_B > 0.1$, even when
shape imperfections and disorder dissolve a shell structure. The
reason is that the number of contributing states (i.e., the states
in the energy interval $\approx [\mu-\Delta_B,\mu+ \Delta_B$) is not
very large for $\delta/\Delta_B > 0.1$. In this case the states
pinned to $\mu$ always make a major contribution to the order
parameter and, so, the profile of the squared absolute value of the
corresponding wave function will mainly determine the spatial
distribution of the condensate. Thus, the domain $\delta/\Delta_B =
0.1$-$1.0$ is in general characterized by strong effects due to the
spatially nonuniform pairing.

We remark that our conclusions do not contradict the usual argument
that spatial variations of the order parameter cost extra energy.
Let us compare a bulk superconductor with a superconducting
nanograin. In bulk the relevant matrix elements controlling the
scattering of the time-reversed states are $-g/V$ and the order
parameter is spatially uniform (in the absence of a magnetic field).
As opposed to bulk, the pair condensate significantly varies with
position in nanograins, which results, of course, in an increase of
the kinetic energy. However, the intrashell matrix elements are now
enhanced in absolute value as compared to $-g/V$ due to quantum
confinement. This compensates energy costs of spatial variations of
the order parameter.

The discussion in the previous paragraph is also related to
arguments that invoke the conventional Ginzburg-Landau theory.
According to this arguments the order parameter is uniform in
samples with size smaller than the bulk coherence length. When
applying this to nanograins, one can conclude that the pair
condensate should not vary with its position. However, this is not
true. It is well-known that one should be careful when applying the
conventional Ginzburg-Landau theory to superconductors with
characteristic size smaller than the zero-temperature (BCS)
coherence length $\xi_0$. Strictly speaking, Gor'kov's derivation of
the conventional Ginzburg-Landau formalism from the BCS approach is
not applicable on a scale smaller than $\xi_0$~(see,.e.g.,
Ref.~\onlinecite{fett}). For ${\rm Sn}$ we have $\xi_0 \approx
230\,{\rm nm}$~(see, e.g., Ref.~\onlinecite{degen}). Thus, in the
case of interest $D \ll \xi_0$, and one can hardly invoke the
conventional Ginzburg-Landau formalism to check whether or not
$\Delta(\rho)$ varies with $\rho$.

\subsection{Confinement-induced Andreev-type states}
\label{sec3c}

Here we would like to discuss one more issue related to a spatially
nonuniform pairing in nanograins. This is the formation of
Andreev-type states induced by quantum confinement~\cite{sh03,sh06}
(see also a similar paper~\cite{torm} discussing Andreev-type states
in an ultracold trapped superfluid Fermi gas). Since the 60s (see
Refs.~\onlinecite{saint,car,andr}) it is known that quasiparticles
can ``feel" a spatial variation of the superconducting order
parameter as a kind of potential barrier. This physical mechanism
(referred to as Andreev mechanism below) is the basis for Andreev
quantization investigated previously for the core of a single vortex
for the mixed state of a type-II superconductor~\cite{car} and for
an isolated normal region of the intermediate state of a type-I
superconductor~\cite{andr} (or for a similar case of SNS
contacts~\cite{saint}). Based on our consideration of
Sec.~\ref{sec3b}, one can expect that Andreev-type states can play a
remarkable role in superconducting nanograins due to significant
spatial variations of the superconducting order parameter. This is
very similar to recently investigated Andreev-type states in
superconducting nanowires/nanofilms~\cite{sh03,sh06}, where the pair
condensate is position dependent in the direction perpendicular to
the nanowire/nanofilm due to the quantization of the perpendicular
electron motion. In Ref.~\onlinecite{sh06} it was shown that
\begin{equation}
\Delta_i =\int\!{\rm d}^3r\,\Delta({\bf r})\,\Bigl[|u_i({\bf
r})|^2+|v_i({\bf r})|^2 \Bigr], \label{andreev}
\end{equation}
which means that the pairing energy gap $\Delta_i$ is the averaged
value of the order parameter "watched" by the quasiparticles with
quantum numbers $i$. Note that $|u_i({\bf r})|^2+|v_i({\bf r})|^2$
can be interpreted as the spatial distribution of quasiparticles
according to the well-known constraint $\int{\rm d}^3r(|u_i({\bf
r})|^2+|v_i({\bf r})|^2)=1$~[see, e.g., Ref.~\onlinecite{ket} and
Eq.~(\ref{constraint})]. When inserting Eqs.~(\ref{factor}) into
Eq.~(\ref{andreev}), one can easily obtain Eq.~(\ref{Delta_matr})
with $\Delta_i=\Delta_{jl}$. If quasiparticles avoid the domains of
enhanced pair condensate, the corresponding integral in the
right-hand-side of Eq.~(\ref{andreev}) becomes smaller and, hence,
such quasiparticles have smaller pairing gaps $\Delta_{jl}$. They
can be referred to as Andreev-type states.

\begin{figure}[tbp]
\resizebox{0.59 \columnwidth}{!}{\rotatebox{0}{
\includegraphics{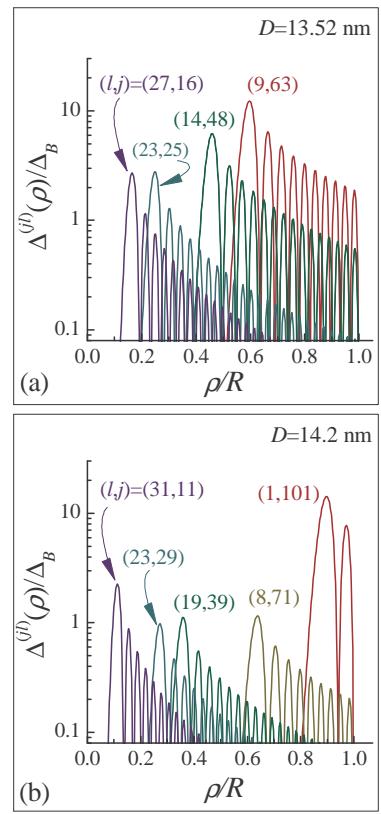}}}
\caption{Shell-dependent contributions to the order parameter
$\Delta_{jl}(\rho)$ for relevant shells: (a), $D=13.52\,{\rm
nm}$, $(j,l)=(27,16),\,(23,25),\,(14,48)$ and $(9,63)$;
(b) $D=14.2\,{\rm nm}$, $(j,l)=(31,11),\,(23,29),\,(19,39),\,
(8,71)$ and $(1,101)$.} \label{fig05}
\end{figure}

Our numerical study of quantum-number dependent pairing gaps
$\Delta_{jl}$ for metallic nanograins reveals a significant role of
Andreev mechanism. Let us consider $D= 13.52\,{\rm nm}$, the
corresponding spatial distribution of the pair condensate is given
in Fig.~\ref{fig03}(b). To show how different species of
quasiparticles are distributed in the radial direction in this case,
the radial-dependent shell contributions (at $T=0$) $\Delta^{(jl)}
(\rho)$~[see Eqs.~(\ref{orderpar}) and (\ref{orderpartial})] are
plotted in Fig.~\ref{fig05}(a). We remark that such a representation
is more informative than simply a plot of $|u_{jl}(\rho)|^2 +
|v_{jl}(\rho)|^2$. First, the radial dependence of $\Delta^{(jl)}
\propto \chi^2_{jl}(\rho)$ is the same as that of $|u_{jl}(\rho)|^2
+|v_{jl}(\rho)|^2 \propto \chi^2_{jl} (\rho)$~[see
Eq.~(\ref{basic})]. Second, a plot of $\Delta^{(jl)} (\rho)$ gives
also information how the corresponding states contribute to
$\Delta(\rho)$. From Fig.~\ref{fig03}(c) we can see that a
significant enhancement of the order parameter occurs at
$\rho/R=0.45$-$0.7$. From Fig.~\ref{fig05}(a) it is clear that this
enhancement is due to the states with $(j,l)= (14,48)$ and $(9,63)$.
Other shells, i.e., $(27,16)$ and $(23,25)$ contribute less, and the
corresponding quasiparticles, representing Andreev-type states, are
mainly located beyond the domain $\rho=0.45-1.0$. As a result, they
have smaller pairing gaps, i.e., $\Delta_{27,16}=2.65\,\Delta_B$ and
$\Delta_{23,25} =2.81\,\Delta_B$, as compared to $\Delta_{14,48}
=4.098\,\Delta_B$ and $\Delta_{9,63}=5.77\,\Delta_B$. As seen, the
quasiparticles with $(j,l)=(27,16)$ are most successful in avoiding
the local enhancement of $\Delta(\rho)$ at $\rho/R= 0.45$-$0.7$ and,
so, $\Delta_{27,16}$ is the smallest pairing gap. Such a
manifestation of Andreev mechanism is not a particular feature of
$D=13.52\,{\rm nm}$. In general, $\Delta_{jl}$ strongly varies with
$j$ and $l$ for diameters $< 30-40\,{\rm nm}$, i.e., where spatial
variations of the order parameter are still pronounced. Quite often
such variations can be an order of magnitude, as, e.g., for
$D=14.2\,{\rm nm}$~(see $\Delta(\rho)$ given in
Fig.~\ref{fig02}(c)). At this diameter a great enhancement of
$\Delta(\rho)$ takes place at $\rho/R=0.9$. This is due to the
contribution of the shell with $(j,l)=(1,101)$~[see
Fig.~\ref{fig05}(b)]. Other shells make much less important inputs
and the corresponding quasiparticles are mainly distributed beyond
the domain $\rho/R =0.9$-$1.0$. So, as compared to $\Delta_{1,101}
=9.32\,\Delta_B$, they have significantly smaller pairing gaps,
i.e., $\Delta_{31,11} = 1.6\Delta_B$, $\Delta_{23,29}
=1.62\,\Delta_B$, $\Delta_{19,39} = 1.72\,\Delta_B$ and
$\Delta_{8,71} = 2.35\,\Delta_B$. Thus, the interplay of Andreev
mechanism and quantum confinement is responsible for variations of
$\Delta_{jl}$ with the relevant quantum numbers.

One could expect that such a serious difference in pairing gaps of
different quasiparticle species can result in a pronounced drop of
the ratio of $\Delta_E$~(the minimal energy gap) to the critical
temperature $k_BT_c$, similar to the case for quantum
superconducting nanowires.~\cite{sh03} The main idea here is that
$\Delta_E$ is governed by Andreev-type states and, hence, is
decreased. Unlike $\Delta_E$, $T_c$ is controlled by the
quasiparticles making a major contribution to $\Delta(\rho)$ and,
so, $T_c$ is coupled to their higher pairing gaps. As a result,
$\Delta_E/k_B T_c$ can be significantly smaller than in bulk. For
instance, one can expect that $\Delta_E=\Delta_{31,
11}=1.6\,\Delta_B$ at $D=14.2\,{\rm nm}$ while $T_c$ is governed by
$\Delta_{1,101} = 9.32\,\Delta_B$. However, this is not correct for
nanograins. The point is that $\Delta_E$ is a spectroscopical gap
which is probed by STM. It is defined as $\Delta_E=\min_{jl}E_{jl}$.
For nanowires the subband-dependent pairing gap is always the
minimal quasiparticle energy due to a quasi-free spectrum in the
direction parallel to the nanowire. For nanograins this is
different. In particular, for $D=14.2\,{\rm nm}$ we have the
following single-electron energies (absorbing $\mu$) of the relevant
shells: $\xi_{31,11}=-18.6\,\Delta_B$, $\xi_{23,29}
=-26.03\,\Delta_B$, $\xi_{19,39}=20.9\,\Delta_B$, $\xi_{8,71}
=22.6\,\Delta_B$ and $\xi_{1,101}=0$. Hence, one can calculate that
$\Delta_E=E_{1,101}=\Delta_{1,101}$ in spite of the fact that
$\Delta_{1,101}$ is the largest pairing gap. Thus, although Andreev
mechanism plays a significant role in superconducting nanograins, it
can hardly be probed by STM-measurements due to the nonzero
interlevel spacing, unlike quantum superconducting nanowires.

\section{Conclusions and discussion}
\label{sec4}

In conclusion, we have shown that the spatial distribution of the
pair condensate is essentially nonuniform in metallic nanograins. In
particular, the spatially nonuniform pairing can proliferate in
nanograins even when $k_FD \sim 300$ and, so, the usual criterion to
neglect variations of the superconducting condensate with position,
i.e., $k_FD \gg 1$, is not very useful and can result in wrong
conclusions. This is the reason why effects due to spatially
nonuniform pairing in superconducting grains were previously
overlooked. Our study suggests that a new criterion should be based
on a more delicate energy scale (as compared to the Fermi energy),
which, in the superconducting state, is given by the bulk order
parameter $\Delta_B$. It turns out that the pairing becomes
spatially nonuniform when the interlevel spacing $\delta$ exceeds
$0.1$-$0.2\,\Delta_B$. Variations of the order parameter with
position exhibit a pronounced enhancement with an increase of
$\delta/\Delta_B$. When $\delta \sim \Delta_B$, such variations can
be almost an order of magnitude in highly symmetric grains. At first
sight, this seems impossible because it costs extra energy for such
spatial variations. However, a nonuniform distribution of the pair
condensate is accompanied by enhanced pairing interaction matrix
elements, which compensates the energy cost for an inhomogeneous
distribution of the condensate. Another point is the size-dependent
pinning of the chemical potential to groups of degenerate or nearly
degenerate energy levels. Such a pinning plays the role of a filter
that increases the contribution of the single-electron levels in the
vicinity of the chemical potential and suppresses contributions of
other states. This results in an additional mechanism favoring
spatially nonuniform pairing in metallic nanograins.

\begin{figure}[tbp]
\resizebox{0.6\columnwidth}{!}{\rotatebox{0}{
\includegraphics{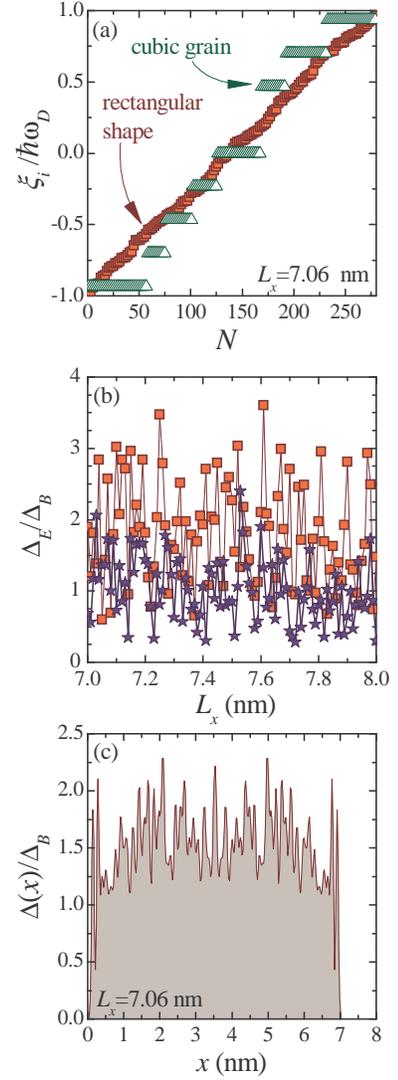}}}
\caption{(a) Single-electron energies $\xi_i$~(given in units of the
Debye energy $\hbar\omega_D$) ordered in ascending manner versus the
ordering number $N$ for the rectangular-shaped aluminum nanograin
with $L_x =7.06\,{\rm nm},\,L_y=L_x/1.1,\,L_z=1.1\,L_x$ (squares)
and for a cubic nanograin with $L_x = 7.06\,{\rm nm}$ (triangles).
(b) The size-dependent excitation gap $\Delta_E/\Delta_B$ versus
$L_x$~(in steps of $\delta L_x= 0.01\,{\rm nm}$) for an aluminum
nanograin of the rectangular shape with the dimensions $L_x,
L_y=L_x/1.1,L_z=1.1\,L_x$: squares represent the results calculated
with the modified matrix elements and with proper variations of
$\mu$; stars are the data obtained for the bulk-like matrix elements
$-g/V$ and $\mu=\mu_B$. (c) The spatial distribution of the pair
condensate in the rectangular grain with $L_x =7.06\, {\rm nm}$ and
$L_y =L_x/1.1,\,L_z=1.1\,L_x$, $\Delta(x)
=\Delta(x,y,z)|_{y=x/1.1,z=1.1x}$.} \label{fig06}
\end{figure}

In this paper we investigated a highly symmetric confining geometry.
Due to this feature the problem becomes effectively one-dimensional
(the order parameter depends only on the radial coordinate) and, so,
sufficiently large diameters up to $D \approx 40\,{\rm nm}$ can be
investigated. This size is almost impossible to reach theoretically
for grains with the order parameter depending on three relevant
coordinates due to time consuming numerical calculations. Such an
effectively one-dimensional problem has large degeneration factors
for the corresponding shell structure, resulting in a significant
enhancement of the pairing correlations. In reality there can be
several issues that may lead to a splitting of the shell levels. It
will decrease the degeneration factors and, so, reduce the pairing
correlations, since the main contribution to the sum in the gap
equation comes from the transitions within the same shell pinned to
the chemical potential. Among such issues is the Jahn-Teller
deformation, i.e., the transformation of a spherical nanograin with
incompletely filled shells to an ellipsoidal shape. In addition, the
surface imperfections and impurities can significantly change the
distribution of single-electron levels. However, our qualitative
results are quite generic and do not depend on a particular shape of
a nanograin and the presence of possible imperfections. For
instance, when $\delta \sim \Delta_B$ the pair condensate will
always be spatially nonuniform because only a few single-electron
levels enters the energy interval $\approx [\mu- \Delta_B,
\mu+\Delta_B]$. Due to the dominant contribution of such levels to
$\Delta({\bf r})$, one can expect that the pair condensate acquires
a profile governed by the squared absolute value of the wave
function for the single-electron state closest in energy to $\mu$.
This is significantly strengthened by an increase (in absolute
value) of the diagonal matrix elements $\langle i,\bar{i}
|\Phi|i,\bar{i} \rangle$ and, in addition, by the pinning of the
chemical potential to the single-particle levels.

We remark that the diagonal matrix elements, i.e., $\langle
i,\bar{i} |\Phi|i,\bar{i}\rangle$~[see the definition for $\Phi$
below Eq.~(\ref{delta_1})] are always enhanced as compared to $-g/V$
in the presence of quantum confinement, whatever disorder and shape
imperfections. This can be seen from the following simple arguments.
Introducing $\varphi_i({\bf r})$, the wave function associated with
state $i$, one can write
$$
\langle i,\bar{i}|\Phi|i,\bar{i}\rangle= -g\int \!\!{\rm d}^3r\;
|\varphi_i({\bf r})|^4.
$$
Due to the normalization condition we have $|\varphi_i({\bf r})|^2
=\frac{1}{V} + d_i({\bf r})$, where $\int{\rm d}^3r d_i({\bf r})=0$.
Then, the above matrix element can be rearranged as
$$
\langle i,\bar{i}|\Phi|i,\bar{i}\rangle= -\frac{g}{V}\Bigl[1+
V\!\int\!\!{\rm d}^3r\,d^{\,2}_i({\bf r})\Bigr].
$$
The second term in the brackets is always positive in the presence
of quantum confinement, i.e., when $d_i({\bf r})\not=0$. It is zero
only when $\varphi_i({\bf r})$'s are chosen in the form of plane
waves, which results in $\langle i,\bar{i}|\Phi|i, \bar{i}
\rangle=-g/V$.

The above discussion can be supplemented by our numerical results
calculated from the BCS-like equation similar to
Eq.~(\ref{selfAnder}) but now for aluminum nanograins of rectangular
shape with dimensions $L_x,\,L_y=L_x/1.1,\,L_z = 1.1L_x$. For
aluminum we have~\cite{degen,fett,ash}: $\hbar\omega_D/k_B=375~{\rm
K}$, $gN(0) = 0.18$, and $\mu_B= 11.67\,{\rm eV}$, which corresponds
to the electron density $n_e=181\,{\rm nm}^{-3}$. In
Fig.~\ref{fig06}(a) single-electron levels arranged in the ascending
order are shown within the Debye window for the rectangular
nanograin with $L_x =7.06\,{\rm nm}$~(squares). The same is also
given here for a cubic aluminum nanograin with $L_x= L_y=
L_z=7.06\,{\rm nm}$~(triangles). As seen, single-electron levels for
the rectangular shape are distributed in a nearly equidistant manner
(with $\delta \approx 0.2$-$0.3\,{\rm meV}\sim \Delta_B$) contrary
to the states corresponding to the cubic geometry. It is well-known
that an almost equidistant distribution~\cite{smit} of
single-electron levels near $\mu$ is also expected in the presence
of significant imperfections such as the surface roughness and/or
impurities. So, our results in Fig.~\ref{fig06} give a feeling about
the role of the spatially nonuniform pairing in disordered metallic
grains. The excitation energy gap $\Delta_E$ for the rectangular
nanograin is shown in units of $\Delta_B$ in Fig.~\ref{fig06} as a
function of $L_x$ in the interval $L_x =7$-$8\,{\rm nm}$. Here
squares represent our results calculated with the modification of
the matrix elements and with $\mu$ varying with $L_x$; stars are the
results found for the bulk-like matrix elements $-g/V$ and $\mu =
\mu_{B}$. As seen, $\Delta_E$~($ \propto T_c$) is now two-times
enhanced as compared to $\Delta_B$~(on average), which is much less
significant than for highly symmetric grains (compare with
Fig.~\ref{fig01}) due to a splitting of the shell levels. However,
the effect of interest is still pronounced: $\Delta_E$ calculated
for the modified matrix elements and with account of size variations
of $\mu$ is generally larger by a factor of $1.5$-$2.0$. The spatial
profile of the order parameter is nonuniform with local enhancements
over its average value by about $100\%$, see, e.g.,
Fig.~\ref{fig06}(c). For rectangular grains with $L_x = 7$-$8\,{\rm
nm}$ we have $\delta \sim \Delta_B$. However, as we checked, the
spatially nonuniform pairing and the related effects of the
modification of the relevant matrix elements and the size-dependent
pinning of $\mu$ are of significance even for smaller $\delta$'s,
i.e., when $\delta > 0.1$-$0.2\, \Delta_B$~($L_x <14$-$15\,{\rm
nm}$). For instance, at $L_x =11\,{\rm nm}$ the order parameter
exhibits variations of about $30$-$40\%$ of its averaged value.
These results are in agreement with our expectations based on the
investigation of the highly symmetric spherical grains.

\begin{acknowledgments}
This work was supported by the Alexander von Humboldt Foundation,
the Flemish Science Foundation (FWO-Vl) and the Belgian Science
Policy (IAP).
\end{acknowledgments}



\begin{thebibliography}{99}
\bibitem{note1} Small corrections to $-\frac{g}{V}$ of about $1\%-3\%$
were considered in Ref.~\onlinecite{gar} for chaotic grains.
\bibitem{note2} Impact of a shell structure on superconducting
correlations were previously considered within the grand canonical
formalism in Refs.~\onlinecite{kres01,kres02} and \onlinecite{gar,bose03}.
However, the authors of Refs.~\cite{kres01,kres02} did not investigate
the dependence of the relevant matrix elements on quantum numbers (and,
so, ignored the spatial dependence of the pair condensate entirely) but
took account of the pinning of the chemical potential to incomplete
shells. In Refs.~\onlinecite{gar,bose03} niether modifications of the
matrix elements due to quantum confinement nor the quantum-size pinning
of the chemical potential were considered.
\end{thebibliography}
\end{document}